\documentclass[useAMS,usenatbib]{mn2e}

\usepackage{graphics,graphicx}

\title[FISICA Observations of NGC 1569]{Probing the Super Star Cluster Environment of NGC 1569 Using FISICA}
\author[D. M. Clark, et al]{D. M. Clark$^{1}$, S. S. Eikenberry$^{2}$, S.~N. Raines$^{2}$, N. Gruel$^{3}$, R. Elston$^{2}$, R. Guzman$^{2}$, \newauthor G. Boreman$^{4}$, P. E. Glenn$^{5}$\\
$^{1}$Instituto de Astronom\'{i}a, Universidad Nacional Aut\'{o}noma
  de M\'{e}xico, Apdo Postal 877, Ensenada, Baja California,
  M\'{e}xico; \\ dmclark@astro.unam.mx \\
$^{2}$Department of Astronomy, University of Florida,
  Gainesville, FL 32611\\
$^{3}$University of Sheffield, Department of Physics and Astronomy, Hicks Building Hounsfield Road Sheffield S3 7RH United Kingdom\\
$^{4}$Dept of Physics and Optical Science University of North Carolina at Charlotte, Charlotte NC 28223 USA\\
$^{5}$Bauer Associates, Inc., 8 Tech Circle, Natick, MA 01760\\}

\begin{document}

\date{in original form 2012 February 25; accepted 2012 October 12}

\pagerange{\pageref{firstpage}--\pageref{lastpage}} \pubyear{2012}

\maketitle

\label{firstpage}

\begin{abstract}
We present near-IR $JH$ spectra of the central regions of the dwarf
  starburst galaxy NGC 1569 using the Florida Image Slicer for
  Infrared Cosmology and Astrophysics (FISICA).  The dust-penetrating
  properties and available spectral features of the near-IR, combined
  with the integral field unit (IFU) capability to take spectra of a field, make FISICA an
  ideal tool for this work.  We use the prominent [He\,{\sc i}]
  (1.083$\mu$m) and Pa$\beta$ (1.282 $\mu$m) lines to probe the dense star
  forming regions as well as characterize the general star forming
  environment around the super star clusters (SSCs) in NGC 1569.  We find [He\,{\sc i}]  coincident with CO clouds to the north and west of the SSCs, which provides the first, conclusive evidence for embedded star clusters here.
\end{abstract}


\section{Introduction}

NGC 1569 is a nearby, 2.2 Mpc \citep{isr90}, dwarf starburst galaxy.
Its centre contains two, prominent super star clusters (SSCs),
designated ``A'' and ``B'' by \citet{dev74}.  Originally thought to be
foreground stars \citep{abl71}, later spectroscopic studies concluded
that they are star clusters associated with NGC 1569
\citep[cf.][]{pra94}.  \citet{hun00} performed a high-resolution, {\it
  HST} optical study of the stellar populations around the SSCs.  They
fit ages of $\geq$7 Myr for A and 10--20 Myr for B.

This galaxy has undergone star formation for almost a complete Hubble
time, containing clusters spanning ages from 4 Myr to 10 Gyr
\citep{alo01,and04,hun00}.  \citet{wal91} suggests this system
underwent six starbursts in its history.  \citet{hun00} find most
clusters are $\la$30 Myr, indicating most cluster formation
occurred at the end of the last starburst episode.  \citet{pas11}
point out that each consecutive starburst has a shorter duration and a
higher star formation rate.  Radio observations show a {$\sim$200 pc diameter hole in the gas
distribution around the SSC A \citep{isr90,alo01}. This indicates a lack of star formation in this region around the SSCs.  This
inner region is also highly disturbed \citep{wes07b} and lacks
supernova remnants (SNRs) \citep{gre02}.  Furthermore, the more massive clusters formed at earlier times, while at later times, only less massive clusters formed \citep{and04}.  \citet{and04} attribute this to the strong, cluster winds of the massive clusters, which prevented large molecular clouds from forming.  

Star formation may still be occurring in a region to the north
\citep{wes07b,tok06}, near cluster 30 as designated by \citet{hun00}
and to the north west, around cluster 10 \citep{hun00}.  The north
west region is quite remarkable.  It is dominated by cluster 10,
designation from \citet{hun00}, which was initially thought to be
another SSC \citep{pra94}, but later evidence suggests it is a less
massive cluster \citep{wes07a,buc00}.  This region itself contains many clusters, with ages of $\sim$25 Myr \citep{and04}.  Just south of 10 lies an
H\,{\sc ii} region \citep{buc00} and to it's west extends a giant
molecular cloud \citep{tay99}.  Using mid-IR observations,
\citet{tok06} find possible evidence for an embedded cluster here,
which could substantiate evidence that star formation is progressing to
the west \citep{gre02}.  \citet{pas11} find that the dust extinction
is higher in the NW than in the SE.

In this article we use the Florida Image Slicer for Infrared Cosmology
and Astrophysics (FISICA), a near-IR image-slicing integral field unit
(IFU), to observe the environment around the SSCs and cluster 10.
Near-IR wavelengths are particularly useful as they contain the
[He\,{\sc i}] line at 1.083 $\mu$m and the Pa$\beta$ line at 1.282
$\mu$m, both of which are diagnostic spectral features of massive
stars and star formation.  In addition, the dust-penetrating
properties of IR wavelengths allows us to probe dense regions hidden
at other wavelengths.  

%Combining these characteristics with the
%ability to take spectra of an entire field at one go, makes FISICA an
%ideal tool for this study.

Our goal in this work is to search for more evidence of massive
star formation in the embedded regions around the SSCs as well as
complement previous work by adding to the understanding of this
complex environment.  We present our work in the following manner: \S2
outlines our observation method and analysis of the data, \S3 presents
a discussion of our results, comparing in detail to past work on this
galaxy, and we summarize our findings in \S4.

\section[]{Observations and Data Analysis}

On October 20, 2004, we acquired $JH$ spectra of NGC 1569 using FISICA
on the KPNO 4-m telescope.  We took five consecutive 300 s
exposures, producing a total on-source exposure of 25 min.  Due to
the extended nature of our target, we moved off-source by
95 arcsec to the SW to measure the sky background.

FISICA \citep{eik06} has a field-of-view (FOV) of 16
by 33 arcsec at f/15.  The IFU divides the field
into 22 strips $\sim$0.7 arcsec wide by 33 arcsec
long.  These strips are rearranged in a line and sent as a
pseudo-longslit through the Florida Multi-object Imaging Near-IR Grism Observational Spectrometer (FLAMINGOS) spectrograph.  All 22 spectra
are imaged on the FLAMINGOS, 2K$\times$2K HgCdTe detector, and each
has a resolution of R $\sim$1300.

We reduced these observations using the data pipeline, Florida Analysis Tool Born Of Yearning for high quality scientific data ({\sc fatboy}).  {\sc fatboy} (C. Warner, et al. 2012, in preparation) calibrates
each spectral frame using flat fields and dark frames and corrects for
cosmic rays, and bad pixels.  All spectra are rectified and all
spectral lines are aligned using skylines.  We did not need to worry
about image distortion as a function of wavelength, because spatial
information is preserved for IFUs that use mirrors.

\begin{figure}
 \centering
\includegraphics[width=40mm]{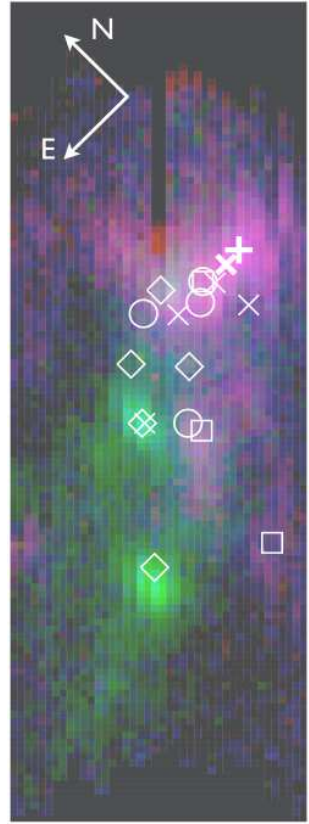}
\caption{False colour image of the central regions of NGC 1569.  North
  is to the upper-left and east is to the lower-left. Green is continuum emission, red is Pa$\beta$ emission and blue is [He\,{\sc i}] emission.  The Pa$\beta$ and [He\,{\sc i}] emission almost coincide spatially.  Diamonds
  designate He\,{\sc ii}  emission peaks, circles [O\,{\sc iii}]
  emission peaks, crosses are [S\,{\sc iv}] emission peaks, and squares
  are positions of H$_2$ regions (see text for details).  The thick,
  white cross is the peak in the  [He\,{\sc i}] emission and the
  thick, white 'X' is the peak in the Pa$\beta$ emission for their
  respective images.  See \S3 for a discussion of the various emission peaks.}
\end{figure}

A useful tool for analyzing IFU data is a data cube.  To make the data
cube, we used a self-written Python program.  The main
difference between a standard image is each pixel has spectral
information.  By this information in the $z$ direction, we produced a
data cube for the FISICA field of NGC 1569.  To ensure Nyquist
sampling, the image of each slice is spread across two pixels.  Considering we had a seeing of $\sim$3.0 pixels, we are Nyquist sampling the Airy disk and this is appropriate.
Therefore, the reproduced image of the field will be foreshortened by
a half in the $x$ direction.  Artificially spreading the flux from each
slice across a width of two pixels, we produced an image with the
correct aspect ratio, having pixels with dimensions of
$\sim$0.3 arcsec.  More specifically, we measured the centroid in flux along each slice.  From this, we measured the fractional flux for either half of the slice.  This fractional flux was then placed in each of the two pixels for the newly binned slice.

\begin{figure*}
\includegraphics[width=140mm]{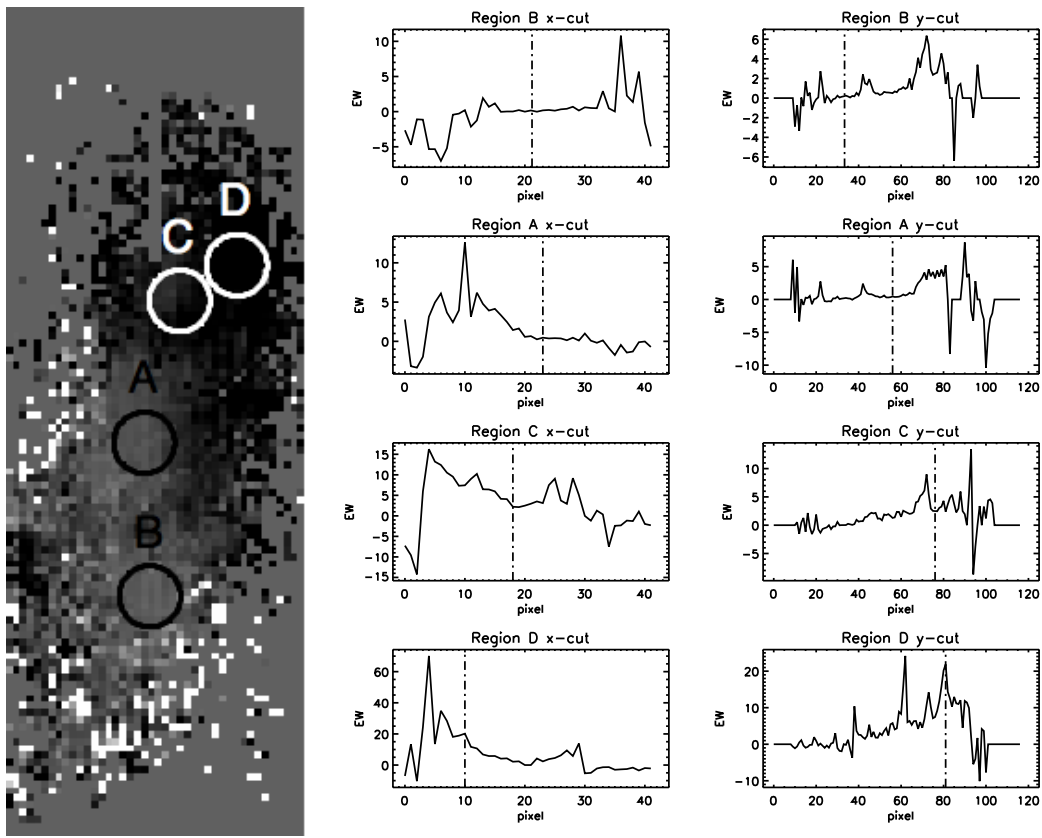}
\caption{[He\,{\sc i}] equivalent width map.  Darker shades
  correspond to regions of higher EW.  Notice the large equivalent
  width corresponds with the bright emission region in the NW and the lower equivalent width around the two SSCs.  The plots on
  the right show vertical and horizontal cuts through each of the
  four regions discussed in \S2.1.1. The regions are labeled as follows: A is SSC A, B is SSC B, C is 
the region around cluster 10, and D is the [He\,{\sc i}] peak.}
\end{figure*}

\subsection{Spectral Maps}

When we initially examined the spectra, we immediately noticed a
prominent [He\,{\sc i}] line in many of the slices.  This type of
emission is often associated with massive stars and star formation.
Considering the importance of this emission, we decided to explore its
extent and relation to the SSCs in NGC 1569.

An easy way to address this query is to make a continuum-subtracted
[He\,{\sc i}] map.  Before making this map, we needed to measure the
width of the [He\,{\sc i}] line and determine the wavelength range
that would include this emission.  First, we summed up all spectra
from all slices, producing a single spectrum for the entire IFU frame.
Fitting a sixth order, Gaussian function to the [He\,{\sc i}] line, we
measured the full width at half maximum (FWHM) and found a value of
FWHM = 10.2 \AA.  We defined an 'extraction width', $w_{ext}$ = 12.8
\AA$ $ (3$\sigma$ of the Gaussian fit).  After summing up all
[He\,{\sc i}] emission within $\pm$$w_{ext}$ of the [He\,{\sc i}] line
centre for each slice, we recombined the slices, producing a
[He\,{\sc i}] map.  We repeated this procedure for two continuum
regions on either side of the [He\,{\sc i}] line, also sampling within $\pm$$w_{ext}$ of each position, producing a pair of continuum
maps.  Averaging the continuum maps, we produced a master,
averaged-continuum map.  Subtracting this from the [He\,{\sc i}] map,
we made the continuum-subtracted [He\,{\sc i}] map.

The continuum-subtracted [He\,{\sc i}] map clearly shows an abundance of
[He\,{\sc i}] at the top, near the set of bright clusters, but this
emission all but disappears for the two SSCs at the middle and bottom
of the frame (see Fig. 1).

A second prominent line apparent in the spectrum is Pa$\beta$ (1.282
$\mu$m).  Like [He\,{\sc i}], Pa$\beta$ is a signature of young,
massive stars, making it a useful diagnostic for star formation in NGC
1569.  Therefore, we also produced a continuum-subtracted map for this
line as well.  Following the same procedure outlined above for
[He\,{\sc i}], we stacked all spectra on the frame and fit a sixth
order Gaussian function to the Pa$\beta$ line.  Measuring a FWHM of
11.3 \AA, we defined $w_{ext}$ = 14.1 \AA (3$\sigma$ of the Gaussian
fit), extracted all Pa$\beta$ emission and produced a Pa$\beta$ map.
Selecting continuum positions on either side of the Pa$\beta$ line, we
extracted continuum emission using $\pm$$w_{ext}$, where $w_{ext}$ is
the same size as that used for the Pa$\beta$ map.  Recombining these
continuum regions and averaging, we produced an averaged-continuum
map.  Finally, subtracting the averaged-continuum map from the
Pa$\beta$ map left us with a continuum-subtracted Pa$\beta$ map.

\begin{table*}
 \centering
 \begin{minipage}{90mm}
	\caption{Flux Fractions for Select Regions of NGC 1569}
\begin{tabular}{@{} lcccc@{}}
\hline
 & Region A & Region B & Region C & Region D \\
 & (SSC A)  & (SSC B) & (Cluster 10 Region) & ([He\,{\sc i}] Pk.) \\
\hline
[He\,{\sc i}]$/$[He\,{\sc i}]$_{tot}$  & 3.36$\times$10$^{-4}$ & 1.03$\times$10$^{-4}$ & 1.36$\times$10$^{-3}$ & 1.79$\times$ 10$^{-3}$\\
 & 2.60$\times$10$^{-5}$ & 2.21$\times$10$^{-5}$ & 3.33$\times$10$^{-5}$ & 6.19$\times$10$^{-5}$\\
Pa$\beta$/Pa$\beta_{tot}$ & 2.27$\times$10$^{-4}$ & 1.31$\times$10$^{-4}$ & 1.13$\times$10$^{-3}$ & 1.58$\times$10$^{-3}$ \\
 & 8.98$\times$10$^{-6}$ & 5.44$\times$10$^{-6}$ & 5.66$\times$10$^{-5}$ & 1.19$\times$10$^{-4}$ \\
Cont.$/$Cont.$_{tot}$ & 2.43$\times$10$^{-3}$ & 3.04$\times$10$^{-3}$ & 1.21$\times$10$^{-3}$ & 4.82$\times$10$^{-4}$ \\
  & 3.22$\times$10$^{-4}$ & 4.84$\times$10$^{-4}$ & 4.06$\times$10$^{-5}$ & 2.81$\times$10$^{-5}$  \\
EW & 0.48 & 0.13 & 4.00 & 18.33 \\
 & 0.34 & 0.18 & 1.08 & 7.87 \\
\hline
\end{tabular}
All ratios and EW are unit less.  Uncertainties are listed below each flux ratio and EW measurement.
\end{minipage}
\end{table*}

Before continuing, we performed a more quantitative analysis to
address how much stronger, or weaker, the [He\,{\sc i}] emission is
compared to the continuum for specific regions in the map.  We chose
regions with a noticeable difference in emission; specifically the two
SSCs, A (centre; region 'A') and B (bottom; region 'B'), the region around cluster 10 (region 'C') and the prominent [HeI] emission region
to it's west, right, (region 'D') (see Fig. 2 and Fig. 3).  Placing
apertures over each region, we measured the total [He\,{\sc i}] flux
in each aperture.  We defined the aperture radius as 4.3 pixels, which
is twice the FWHM of the Gaussian point-spread-function of the SSCs.
This choice in aperture size was sufficient to encompass each SSC
without a substantial contribution from background sources.  Dividing
the fluxes by the total flux for the entire [He\,{\sc i}]
continuum-subtracted map, gave us the fraction of [He\,{\sc i}]
emission each region contributes to the IFU frame (see Table 1).
Repeating this process with the same-sized apertures for the
averaged-continuum map gave us the fraction of continuum emission each
region contributes to the IFU frame (see Table 1).  Notice the
increase in intensity of [He\,{\sc i}] flux from bottom-to-top, region
'B' to 'A' to 'C' to 'D', while the opposite effect is seen
for the continuum flux.

Repeating the same quantitative analysis on the Pa$\beta$
map that we performed on the [He\,{\sc i}] map, we computed the flux
ratio in Pa$\beta$ flux for regions A, B, C and D to total Pa$\beta$
flux for the image.  Lastly, using
both continuum-subtracted maps, [He\,{\sc i}] and Pa$\beta$, plus an
averaged continuum map, we produced a false colour image,
high-lighting the various emission regions (Fig. 1).

We obtained further information on the central parts of NGC 1569 by
making an equivalent width (EW) map.  This is simply the [He\,{\sc i}]
continuum-subtracted map divided by the averaged-continuum map.
Examining this EW map shown in Fig. 2, shows a region of large EW
in the NW as well as a second peak to the north of cluster A.  The
highest EW is in the NW region and has a value of $\sim$64 (See Fig. 2).  The peak
in EW to the north is about a factor of ten lower in EW.  Also, notice
the much lower EW at the positions of the two SSCs. Next to the EW
map we show plots of vertical and horizontal cuts through the centre
of each region.   Each cut is the median average of $\pm$2 pixels from the line.

\subsection{Spectra}

In addition to spectral maps, we also made spectra from the four regions A, B, C and D defined in \S2.1 (see Fig. 3).  These spectra were taken from the centre of each region and were produced by median combining the spectra $\pm$2 pixels from the centre in the $y$-direction.  Notice that the spectra for SSC B (Region B) is dominated by the continuum, while the spectra for the [He\,{\sc i}] peak (Region D) is dominated by the emission lines [He\,{\sc i}] and Pa$\beta$.  

\begin{figure}
\includegraphics[width=90mm]{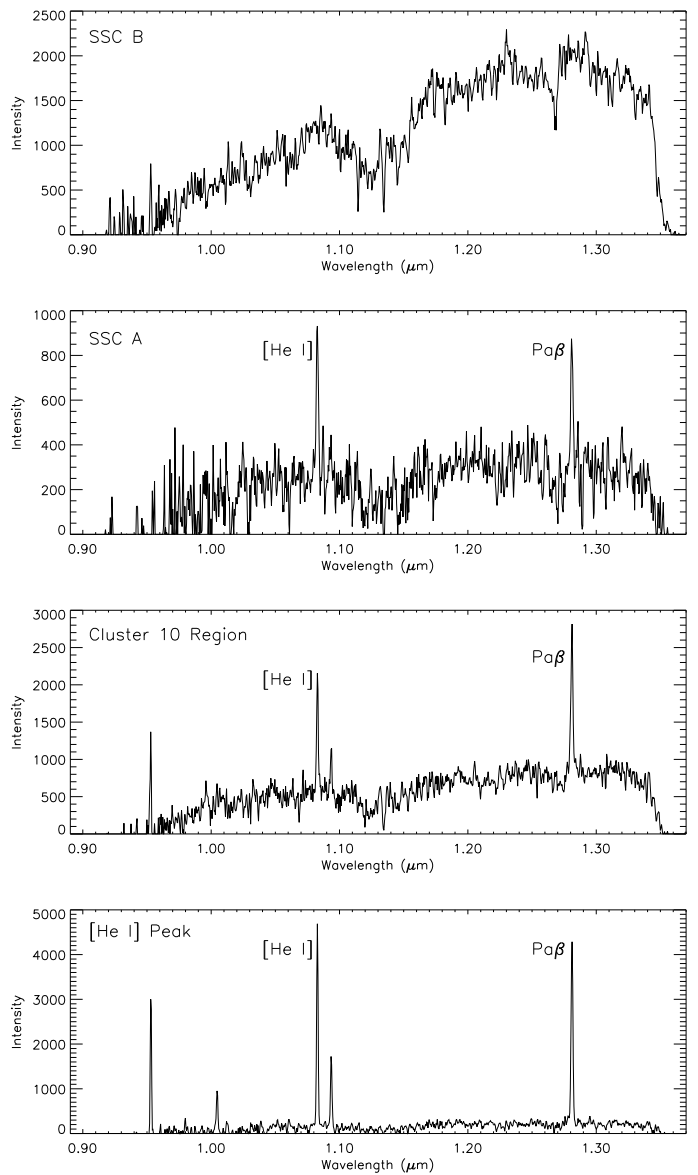}
\caption{Spectra through each region as labeled in Figure 2.  Notice that the continuum emission is dominant in the spectra for SSC B, but the emission lines [He\,{\sc i}] and Pa$\beta$ are dominant near the [He\,{\sc i}] peak.}
\end{figure}

\section{Discussion}

Thus far we have presented a description of the [He\,{\sc i}] and
Pa$\beta$ emission in the central regions of NGC 1569.  We now proceed
by comparing our observations to other studies at a variety of
wavelengths to try to increase understanding of this complex region.

As mentioned above, there is a distinct lack of both [He\,{\sc i}] and
Pa$\beta$ emission present in both SSCs.  While SSC A may contain
trace amounts of these emission lines, SSC B is clearly devoid of any
emission at these wavelengths.  He\,{\sc ii}  observations also show a
similar trend in emission properties \citep{buc00}.  These observations are in agreement with H I observations, which indicate a ~200 pc hole in the H I distribution, centered on SSC A \citep{isr90,sti98,alo01}.  This suggests that star formation has ceased in the inner regions of NGC 1569 \citep{alo01}.  A detailed
optical study of the stellar populations in these clusters indicates A
contains luminous blue variables (LBVs), while B is dominated by red supergiants \citep{hun00}.  Thus the [He\,{\sc i}] emission is most likely from the hot,
blue stars in A, while the more evolved, massive stars in B explains
why we do not see any emission in this cluster.

To the north of SSC B lies a small clump of [He\,{\sc i}] and
Pa$\beta$ emission.  This region exhibits a relatively large equivalent width compared to the field,
peaking at $\sim$23.  It is south of cluster 30, as designated by
\citet{hun00}, which is off our field.  As seen in Fig. 1, there is
no continuum emission, which is consistent with $V$-band images, where
it is only faintly detected \citep{hun00}, as well as recent optical
IFU observations \citep{wes07b}.  This contrasts with narrow-band
[O\,{\sc iii}] and H$\alpha$ images which show prominent emission here
\citep{buc06}.  There is also a non-thermal radio source at this
position \citep{gre02}.  Furthermore, CO emission is observed at this
location \citep{tay99}, possibly indicating this source is an
embedded, young star cluster, as suggested by \citet{wes07b}.  In
their detailed study of this area, \citet{wes07b} were not able to
detect any stellar signatures.  Thus, our detection of [He\,{\sc i}]
is the first proof that this source is a young cluster forming
from a blister in an expanding bubble of gas \citep{wes07b}.

Just south of SSC A appears another region of emission.  It is most
prominent in [He\,{\sc i}], but also seen in Pa$\beta$ and continuum.
This source is a known H\,{\sc ii} region \citep{wal91} and is also seen in
[O\,{\sc iii}] images \citep{buc06}.  In \citet{wes07b}, there is
evidence for an unusual clump with a high velocity and large line
widths.  These authors suggest this clump lies along our sight-line and
is not necessarily connected with the central regions of the galaxy.
In our equivalent width map, we do not see evidence for large line
widths here, but then we lack the spatial resolution of GMOS.

One of the most intriguing parts of NGC 1569 sits NW of SSC A, where star formation still appears to be going on. Optical images show an abundance of star clusters and bright stars
here \citep{hun00, and04}.  Many of the clusters in this region have ages of $\sim$25 Myr \citep{and04}. The region is dominated by cluster 10 as
designated by \citet{hun00}.  Narrow-band, He\,{\sc ii}  observations
show emission coincident with the cluster, and the authors hypothesize
this emission is due to three WNL stars or is nebular in origin
\citep{buc00}.  Recent, optical luminosity measurements of cluster 10
confirm that massive stars are the best explanation for this emission
\citep{wes07a}.  \citet{wes07a} acquired {\it HST} images of this cluster and show that it is actually two clusters, 10A and 10B.  Using photometry, they estimate 10A is between 5-7 Myr, while 10B is $<$5 Myr.  In addition, cluster 6 (as designated by \citet{hun00}) also has a young age, ~4 Myr \citep{and04}.  These cluster ages indicate a younger population of clusters here.  Furthermore, \citet{pas11} find evidence for a higher star formation rate in the NW.

South of cluster 10, and on the edge of this large [He\,{\sc i}]
complex, sits the H\,{\sc ii} region identified by \citet{wal91}.  This
nebular region is coincident with [O\,{\sc iii}] \citep{buc06} and
[S\,{\sc iv} ] \citep{tok06} emission.  It also sits perched on the edge
of a giant molecular cloud (GMC) seen in CO observations \citep{tay99}.

Progressing farther south we find the strongest peak in both
[He\,{\sc i}] and Pa$\beta$.  The Pa$\beta$ peak is offset to the
north of the [He\,{\sc i}] peak by 0.6 arcsec (see Fig. 1).  Considering the average shifts in the image slices used to create the reconstructed image is 0.3$\pm$0.8 arcsec, we can not reliably differentiate the positions of the [He\,{\sc i}] and Pa$\beta$ peaks.

Interestingly, the
[He\,{\sc i}] peak corresponds with the brightest mid-IR, [S\,{\sc iv}] peak observed by \citet{tok06}.  These authors suggest this
is an embedded cluster consisting of 40 O7 stars.  Our observations
of coincident [He\,{\sc i}] emission seems to confirm their finding
and is the first conclusive evidence that a cluster is forming
here.  Furthermore, comparing the location of the [He\,{\sc i}] and
[S\,{\sc iv}] peaks to CO maps presented in \citep{tay99}, suggest this
cluster is buried in the GMC that extends to the southwest.

This discussion demonstrates how newly formed star clusters provoke the continued formation of star clusters in the local environment of NGC 1569.  \citet{and04} mention that star formation is occurring along bubble walls, indicating self-propagating star formation, which is discussed by several authors.  \citet{buc06} find that the surface density of H\,{\sc ii} regions remains constant with in annuli as the annuli are moved away from the center.  They suggest this indicates feedback effects are confined to the center of the galaxy on scales of 1 pc.  \citet{wes07b} use {\it HST} observations to show the cometary appearance of molecular clouds, indicating the effects of stellar winds from the inner clusters.  Lastly, \citet{pas11} find that line ratios of [Fe\,{\sc ii}] 1.64$\mu$m / Br$\gamma$ are highest at the edge of gas-filled cavities and in the outskirts of the galaxy.  This indicates the average between past and on-going star formation.  In addition, it indicates that star formation has been quenched in the interior and more recently, triggered in a ring around the cavities $\sim$4 Myr ago, after the SSCs formed.  Our observations of [He\,{\sc i}]  coincident with CO clouds provide evidence for embedded clusters that were triggered to form by cluster winds near the CO clouds.

\section{Conclusions}

Our observations of clumped [He\,{\sc i}] emission to the N of SSC A and in the NW provide the first, conclusive evidence of newly forming star clusters here.  Combining our observations, with additional, multi-wavelength data, we discussed a scenario of the on-going star formation in this galaxy.

\section*{Acknowledgments}

DMC and SSE were supported in part by a NSF award (NSF-0507547) for
this work.  RG and NG acknowledge funding from NASA-LTSA grant
 NAG5-11635.  Additional support for DMC was provided by a UNAM
postdoctoral fellowship. FISICA was funded by the UCF-UF Space
Research Initiative.

\label{lastpage}


\begin{thebibliography}{99}

\bibitem[\protect\citeauthoryear{Ables}{1971}]{abl71} Ables, H.D., 1971, Publ. U.S. Naval
  Obs. XX, Part IV, p.60

\bibitem[\protect\citeauthoryear{Aloisi et al.}{2001}]{alo01} Aloisi, A., et al.  2001, AJ,
  121, 1425

\bibitem[\protect\citeauthoryear{Anders et al.}{2004}]{and04} Anders, P., de Grijs, R.,
  Fritze-v. Alvensleben, U., \& Bissantz, N.  2004, MNRAS, 347, 17

\bibitem[\protect\citeauthoryear{Bisbas et al.}{2011}]{bis11} Bisbas, T.~G.,  W{\"u}nsch, R., Whitworth, A.~P., Hubber, D.~A.,  \& Walch, S.\ 2011, ApJ, 736, 142

\bibitem[Brand et al.(2011)]{bra11} Brand, J., Massi, F., Zavagno, A., Deharveng, L., \& Lefloch, B.\ 2011,A\&A, 527, A62

\bibitem[\protect\citeauthoryear{Buckalew et al.}{2000}]{buc00} Buckalew, B.A., Dufour, R.J.,
  Shopbell, P.L., \& Walter, D.K.  2000, AJ, 120, 2402

\bibitem[\protect\citeauthoryear{Buckalew \& Kobulnicky}{2006}]{buc06} Buckalew, B.A. \&
  Kobulnicky, H.A.  2006, AJ, 132, 1061

\bibitem[\protect\citeauthoryear{Eikenberry et al.}{2006}]{eik06} Eikenberry, S., et al.\ 2006, ProcSPIE, 6269, 146

\bibitem[\protect\citeauthoryear{de Vaucouleurs et al.}{1974}]{dev74} de Vaucouleurs G., de
  Vaucouleurs A., Pence W., 1974, ApJ, 194, L119

\bibitem[\protect\citeauthoryear{Greve et al.}{1996}]{gre96} Greve, A., Becker, R., Johansson,
  L.E.B., \& McKeith, C.D.  1996, A\&A, 312, 391

\bibitem[\protect\citeauthoryear{Greve et al.}{2002}]{gre02} Greve, A., Tarchi, A.,
  Huttemeister, S., De Grijs, R., Van Der Hulst, J.M., Garrington, S.,
  \& Neininger, N.  2002, A\&A, 381, 825

\bibitem[\protect\citeauthoryear{Hunter et al.}{2000}]{hun00} Hunter, D., O'Connell, R.W.,
  Gallagher, J.S., \& Smecker-Hane, T.A.  2000, AJ, 120, 2383

\bibitem[\protect\citeauthoryear{Israel \& van Driel}{1990}]{isr90} Israel, F.P. \& Van Driel,
  W.  1990, A\&A, 236, 323.

\bibitem[\protect\citeauthoryear{Koenig et al.}{2012}]{koe12} Koenig, X.~P.,  Leisawitz, D.~T., Benford, D.~J., et al.\ 2012, ApJ, 744, 130

\bibitem[\protect\citeauthoryear{Lisenfeld et al.}{2005}]{lis05} Lisenfeld, U., 
Israel, F.~P., Stil, J.~M., Sievers, A., \& Haas, M.\ 2005, The Spectral 
Energy Distributions of Gas-Rich Galaxies: Confronting Models with Data, 
761, 239 

\bibitem[\protect\citeauthoryear{Pasquali et al.}{2011}]{pas11} Pasquali, A., et al.\ 2011,
  AJ, 141, 132

\bibitem[\protect\citeauthoryear{Prada}{1994}]{pra94} Prada, F., Greve, A., \& McKeith C.D.
  1994, A\&A, 288, 396

\bibitem[\protect\citeauthoryear{Stil \& Israel}{1998}]{sti98} Stil, J.~M., \& Israel, F.~P.\ 1998, A\&A, 337, 64

\bibitem[\protect\citeauthoryear{Taylor et al.}{1999}]{tay99} Taylor, C.L., Huttemeister, S.,
  Klein, U., \& Greve, A.  1999, A\&A, 349, 424

\bibitem[\protect\citeauthoryear{Tokura et al.}{2006}]{tok06} Tokura, D., et al.  2006, ApJ,
  648, 355

\bibitem[\protect\citeauthoryear{Waller}{1991}]{wal91} Waller, W.H.  1991, ApJ, 370, 144

\bibitem[\protect\citeauthoryear{Westmoquette et al.}{2007a}]{wes07a} Westmoquette, M.~S., Exter,
  K.~M., Smith, L.~J., \& Gallagher, J.~S.\ 2007a, MNRAS, 381, 894

\bibitem[\protect\citeauthoryear{Westmoquette et al.}{2007b}]{wes07b} Westmoquette, M.~S., Smith,
  L.~J., Gallagher, J.~S., \& Exter, K.~M.\ 2007b, MNRAS, 381, 913\end{thebibliography}
\end{document}